\documentclass[a4paper, amsfonts, amssymb, amsmath, reprint, showkeys, nofootinbib, superscriptaddress, twoside, aps, pra]{revtex4-1}

\usepackage{bm}
\usepackage{tikz}
\usepackage{hyperref}
\usepackage{physics}
\usepackage{amsmath}

\usepackage[T1]{fontenc} 
\usepackage{lmodern}

\usepackage{etoolbox} 
\apptocmd{\thebibliography}{\raggedright}{}{}

\newcommand\figref{FIG.~\ref}		
\newcommand\equref{eq.~\eqref}		
\newcommand\appref{Appendix~\ref}	
\newcommand\secref{Section~\ref}

\begin{document}

\title{Quasi-One-Dimensional Few-Body Systems with Correlated Gaussians}%

\author{M. Wallenius}%
\affiliation{Department of Physics and Astronomy, Aarhus University, DK-8000 Aarhus C, Denmark.}%

\author{D. V. Fedorov}%
\affiliation{Department of Physics and Astronomy, Aarhus University, DK-8000 Aarhus C, Denmark.}%

\author{A. S. Jensen}%
\affiliation{Department of Physics and Astronomy, Aarhus University, DK-8000 Aarhus C, Denmark.}%

\author{N. T. Zinner}%
\email{Zinner@aias.dk}%
\affiliation{Department of Physics and Astronomy, Aarhus University, DK-8000 Aarhus C, Denmark.}%
\affiliation{Aarhus Institute of Advanced Studies, Aarhus University, DK-8000 Aarhus C, Denmark.}%

\date{\today}

\begin{abstract}
The theoretical study of ultracold few-body systems is often done using an idealized 1D model with zero range interactions. Here we study these systems using a more realistic 3D model with finite range interactions. We place three-particles, two identical and one impurity, in an axial symmetric harmonic trap and solve the corresponding stationary Schr\"odinger equation using the correlated Gaussian method for different particle types, aspect ratios and interactions strength. We show that the idealized model is accurate for small and intermediate strength interactions at aspect ratios larger than four, independently of the particle types. In the strongly interacting limit, the idealized model is acceptable for bosonic systems, but not for fermionic systems even at large aspect ratios.
\end{abstract}

\maketitle

\section{Introduction}
The field of ultracold one-dimensional (1D) few-body systems has received a lot of attention since Bethe in $1931$ published his exact solution to the 1D Heisenberg model \cite{Bethe1931}. 1D systems are of particular interest due to their simplicity and different properties compared to the two-dimensional (2D) and three-dimensional (3D) systems. For example, in 1D particles have to go through each other, and therefore interact, in order to exchange positions. The realization of 1D systems \cite{bongs2001waveguide, schreck2001quasipure, gorlitz2001realization, dettmer2001observation, moritz2003exciting, Pangano2014onedimensional} made it possible to verify theoretical models, such as the Tonks-Girardeau gas \cite{tonks1936complete, paredes2004tonks, kinoshita2004observation} and simulate quantum magnetism \cite{lindgren2014fermionization, Sowi_ski_2019, PhysRevA.90.013611, Deuretzbache2017tuning, Deuretzbacher2017spin, PhysRevLett.111.045302, murmann2015antiferromagnetic, LiapSpinImbalance2010, Dehkharghani2014quantum, CuiGroundState2013}. Other systems such as organic conductors \cite{jerome1980superconductivity} and nanowires \cite{arutyunov2008superconductivity} are of 1D nature as well.

1D systems are formed in highly controllable environments where the particle number, interaction strength, internal states and motional states can be controlled \cite{paredes2004tonks, Kinoshita2005, wenz2013few, haller2009realization, serwane2011deterministic}. In particular, it is possible to experiment with mixtures of particles with different masses \cite{spethmann2012inserting, wenz2013few}, while a system of identical particles can be made distinguishable utilizing different hyperfine or spin states. The interaction strength is controllable through Feshbach resonance \cite{bloch2008many} while for systems under external confinement it also depends on the geometry of the trap \cite{olshanii1998atomic}.

1D systems are often studied using an idealized model assuming an exact 1D trap and zero-range interactions under the assumption that this reproduces the same physics as the real world quasi-1D systems. Experimentally such systems are formed in anisotropic harmonic traps where the motional degrees of freedom are cooled down below the transverse excitation energy, thus effectively generating a 1D system. The axially symmetric harmonic trap is parameterized by the frequencies in the transverse, $\omega_\perp$, and longitudinal, $\omega_z$, directions. For example \cite{zurn2012fermionization,murmann2015antiferromagnetic, PhysRevA.90.013611} used an aspect ratio, $\eta\equiv \tfrac{\omega_\perp}{\omega_z}\sim 10$, while \cite{kinoshita2004observation} used an extreme value of $\eta=350$. The transition by such confinement from 3D to lower dimensions is in this way continuous. This suggests that a given anisotropy corresponds to a non-integer dimension as in \cite{gar19} and \cite{chr18} for two and three particles. However, it is not known which confinement is required to make 1D (or 2D) models a good approximation, how one can quantify the relevant criteria, what the decisive quantities are and when the specific properties are reproduced.

The purpose of this paper is to address these questions. We want to lay the ground for treatments of interacting particles in anisotropic harmonic external traps. We shall at first restrict ourselves to investigations of the simplest systems. In general, broadly applicable results can only be expected for short-range (yet finite) interacting particles, since then the dominating parts of the wave function occur when the particles are outside the range of their potentials. Otherwise, the results could too easily depend on the individual potentials. Hence we restrict ourselves to short-range interactions and ground states. We study the system by solving the corresponding stationary Sch\"odinger equation using the correlated Gaussian method. This returns the energies and wave functions from which all necessary information can be obtained.

Exact solutions exist for systems of two short-range interacting particles in exact dimensionalities \cite{busch1998two,farrell2009universality, koscik2018exactly, koscik2019exactly} and in an axial symmetric trap \cite{idziaszek2005, idziaszek2006analytical}. To establish the validity of our methods, we shall therefore first investigate these simple quasi-1D systems and compare with the results from these exact two-body calculations. For the three-particle systems, no exact solution is known for interactions of finite range or strength, and numerical methods are unavoidable. For infinite interaction strength, exact solutions are known for any number of particles and trap potential \cite{volosniev2014strongly, Volosniev2014}. \\\hspace*{\fill}

The paper is organized as follows. In \secref{sec:System_and_method} we present the system, necessary theory and the method. We test the method in \secref{sec:1+1} and the results are presented and discussed in \secref{sec:2+1}. We end with a conclusion and outlook in \secref{sec:Con_and_out}.

\section{System and Method}\label{sec:System_and_method}
In this section, we introduce the systems of interest and the corresponding Hamiltonian. As we seek to compare the 3D solution with the idealized 1D model we introduce the effective 1D interaction, $g_\text{1D}$, in terms of 3D parameters. At last, to solve the corresponding Schr\"odinger equation we introduce a set of Jacobi coordinates and the correlated Gaussian method.
\subsection{System}
The Hamiltonian for a general harmonically trapped system consisting of $N$ particles with mass $m_i$ and positions $\vec{r}_i$, for $i\in \{1,2,...,N\}$, is
\begin{align}\label{eq:N-body_Hamiltonian}
\begin{split}
	\hat{H}&=\sum_{i=1}^N\left[-\frac{\hbar^2}{2m_i}\nabla^2_{\vec{r}_i}+\frac{1}{2}m_i\Big(\omega_x^2 x_i^2+\omega_y^2 y_i^2+\omega_z^2 z_i^2\Big)\right]\\
	&+\sum_{j>i=1}^{N} S_{ij}e^{-\left(\vec{r}_i-\vec{r}_j\right)^2/r_0^2}, 
\end{split}
\end{align} 
where $\hbar$ is Planck's reduced constant, $\omega_c$, for $c=x,y,z$, is the trap frequencies, $\nabla^2_{\vec{r}_i}$ is the Laplace operator for the $i$'th particle, $S_{ij}$ is the Gaussian interaction strength between the $i$'th and $j$'th particle and $r_0$ its range. As axial symmetric systems will be considered we parameterize the external trap in terms of $\omega_x=\omega_y=\omega_\perp$ and $\omega_z$. Further, we could have chosen any desired interaction potential as long as its range is much smaller than the oscillator length \cite{farrell2009universality}. We choose a Gaussian as this is easily implemented in the correlated Gaussian method. In addition, only two-body interactions are considered as three-body and more are negligible due to the dilluteness of the systems.

We will focus on systems of two identical particles called the majority, either fermions or bosons, interacting with a distinguishable particle called the impurity. These systems will be denoted 2f+1 and 2b+1 respectively, where the former has been studied experimentally in \cite{murmann2015antiferromagnetic}. For the fermionic (bosonic) system the majority particles will be labeled $\uparrow$ ($A$) while the impurity $\downarrow$ ($B$).
\begin{figure}[ht]
	\centering
	\begin{tikzpicture}
	\node[inner sep=0pt] (whitehead) at (0,0)
	{\includegraphics[]{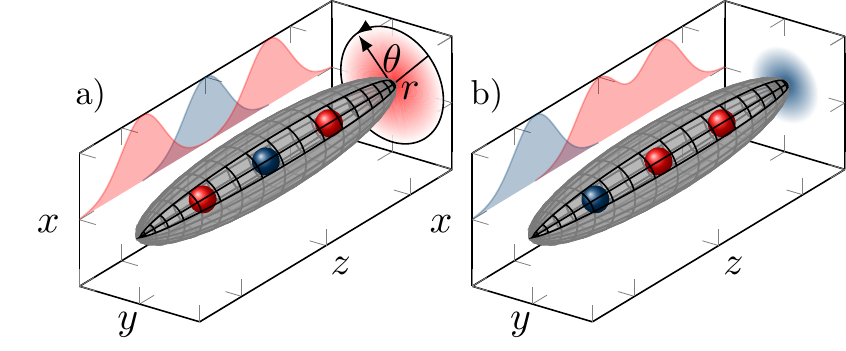}};
	\end{tikzpicture}
	\caption{\label{fig:2+1_system}Visualization of the 2+1 system trapped in an axial symmetric harmonic trap. The system consists of two identical particles, red, and one impurity, blue. The plots in the $xz$-pane show an example of the density distribution. Panel (a) show the particle configuration $AAB$ ($\uparrow\downarrow\uparrow$) and panel (b) the configuration $AAB$ ($\uparrow \uparrow \downarrow$) for bosons (fermions).}
\end{figure}
The two possible configurations of the 2+1 system can be seen in \figref{fig:2+1_system}. Here two majority particles (red) and the impurity (blue) are trapped in an axial symmetric harmonic trap. The density distribution in the $z$-axis is shown in the $xz$-plane. As \equref{eq:N-body_Hamiltonian} is parity invariant the energy eigenstates are also parity eigenstates. 

We will assume that the inter-species interactions are much stronger than the intra-spices which is therefore neglected. Further, we will only consider mass-balanced systems, even though the formalism introduced in the following sections is completely general.

\subsection{Interactions under External Confinement}
We relate the Gaussian interaction strength $S_{ij}$ from \equref{eq:Hamiltonian_system} to the s-wave scattering length $a_\text{3D}$. Further, in 1D systems the interaction can be modeled using a Dirac $\delta$-function as $V_\text{1D}(x)=g_\text{1D}\delta(x)$ where,
\begin{align}\label{eq:g_1D}
g_\text{1D}=-\frac{\hbar^2}{\mu a_\text{1D}},
\end{align}
is the 1D coupling constant \cite{olshanii1998atomic}. The quantity $a_\text{1D}$ is known as the 1D scattering length and is given in terms of $a_\text{3D}$ as, 
\begin{align}\label{eq:a_1D}
a_\text{1D}=-\frac{a^2_\perp}{2 a_\text{3D}}\left(1-C \frac{a_\text{3D}}{a_\perp}\right).
\end{align}
Here $a_\perp=\sqrt{\frac{\hbar}{\mu \omega_z}}$ is the length scale of the transverse trap and $C=-\zeta\left(\frac{1}{2}\right)\approx 1.460$ where $\zeta$ is the Riemann zeta function. Using \equref{eq:g_1D} and (\ref{eq:a_1D}) we can relate $S_{ij}$ to the effective 1D interaction $g_\text{1D}$. Notice that if the scattering length approaches the length scale of the transverse trap a resonance occurs. This is known as confinement-induced resonance (CIR). CIR occurs when the energy of two scattering atoms becomes degenerate with a transverse excited molecular bound state \cite{zhang2011confinement}. It has been confirmed both numerically and experimentally \cite{paredes2004tonks, haller2009realization, kinoshita2004observation, haller2010confinement, zurn2012fermionization}. CIR is crucial when studying quasi-1D systems as it allows for $g_\text{1D}$ to be tuned from strongly repulsive to strongly attractive through the geometry of the trap. In the following, all results will be expressed in terms of $g_\text{1D}$.

\subsection{Jacobi Coordinates}
The complexity of solving the Schr\"odinger equation can effectively be reduced utilizing a set of relative coordinates. We choose a standard set of Jacobi coordinates, $\bm x=(x_1, x_2,..., x_{3(N-1)})^\text{T}$ for the relative motion along with the center of mass, $\vec{x}_N=(x_{3N-2}, x_{3N-1}, x_{3N})^\text{T}$. These are related to the particle coordinates as $\tilde{\bm x}=U\bm r$ where $\bm r=(r_1, r_2,...,r_{3N})^\text{T}$ and $\tilde{\bm x}^\text{T}=(\bm x^\text{T}, \vec{x}_N^\text{T})$. The Jacobi transformation matrix $U$ is defined as 
\begin{align}\label{eq:Jacobi-Matrix}
	U=\begin{pmatrix}
	1 & -1 & 0 & \cdots& 0\\ 
	\frac{m_1}{M_2} & \frac{m_2}{M_2} & -1 &\cdots  & 0\\ 
	\vdots & \vdots & \vdots & \ddots & \vdots \\ 
	\frac{m_1}{M_{N-1}} & \frac{m_2}{M_{N-1}}  & \frac{m_3}{M_{N-1}}  & \cdots & -1\\ 
	\frac{m_1}{M_{N}} & \frac{m_2}{M_{N}} & \frac{m_3}{M_{N}} & \cdots & \frac{m_N}{M_{N}}
	\end{pmatrix}\otimes I_{3\cross 3}
\end{align}
where $M_n=\sum_{i=1}^{n}m_i$ and $I_{3\cross 3}$ is the $3\cross 3$ identity matrix. An analogous transformation can be performed on the gradient operator as $\grad_{\bm r} =U^\text{T} \grad_{\tilde{\bm x}}$ where $\grad_{\bm c}=(\pdv{c_1}, \pdv{c_2},...,\pdv{c_{3N}})^\text{T}$ where $c=r, x$. Expressing the system in terms of relative coordinates is not always possible when an external potential is present but, as all particles experience the same trap, the considered systems can be. In Jacobi coordinates and in harmonic oscillator units where length and energy are measured in terms of $b\equiv \sqrt{\tfrac{\hbar}{\mu \omega_z}}$ and $\hbar \omega_z$, the Hamiltonian in \equref{eq:N-body_Hamiltonian} is given as
\begin{align}\label{eq:Hamiltonian_system}
\begin{split}
	\hat{H}=&-\frac{1}{2}\grad_{\bm x}^\text{T}\Lambda \grad_{\bm x}+\bm x^\text{T} \Omega \bm x+\sum_{j>i=1}^{N}S_{ij}e^{- \bm x^\text{T} W_{ij} \bm x/r_0^2}\\
	&+\hat{H}_\text{CM},
\end{split}
\end{align}
where 
\begin{align}
	\Lambda\equiv&   \text{diag} \left(\frac{\mu}{\mu_1}, \frac{\mu}{\mu_2},...,\frac{\mu}{\mu_{N-1}}\right) \otimes I_{3\cross 3},\\
	\Omega \equiv& \frac{1}{2} \Lambda^{-1} \otimes \text{diag}\left(\eta^2, \eta^2, 1\right),
\end{align}
$\mu$ is an arbitrary mass scale and $\mu_n\equiv\tfrac{M_n m_{n+1}}{M_{n+1}}$. The matrix $W_{ij}$ is here defined through the operation $\bm x^\text{T} W_{ij} \bm x=\sum_{c=x,y,z}(r_{i,c}-r_{j,c})^2$ (see \appref{app:matrix-elements} for more information). As this transformation leads to a separable Hamiltonian the wave function can be written as a product of the relative and the center-of-mass wave functions as $\Psi({\tilde{\bm x}})=\psi(\bm x) \phi(\vec{x}_N)$. The center-of-mass Hamiltonian is given as 
\begin{align}
	\hat{H}_\text{CM}=-\frac{\mu}{2M_N}\nabla^2_{\vec{x}_N}+\vec{x}^\text{T}_N \Omega_\text{CM} \vec{x},
\end{align} 
where $\Omega_\text{CM}\equiv \tfrac{1}{2} \tfrac{M_N}{\mu} \text{diag}(\eta^2, \eta^2,1)$ from which the ground state solution follows naturally as
\begin{align}\label{eq:CM-solution}
	\phi(\vec{x}_N)=\sqrt{\eta}\left(\frac{M_N}{\pi \mu}\right)^{\frac{3}{4}}e^{-\frac{M_N}{2\mu}\left(\eta(x_N^2+y_N^2)+z_N^2\right)}.
\end{align}
We have now effectively reduced a system of $N$-particles to an $(N-1)$-particle problem. Left is to solve the Schr\"odinger equation with the relative Hamiltonian. We do this using the correlated Gaussian method.

\subsection{Correlated Gaussian Mehod}
The correlated Gaussian method (CGM) is a popular variational method used to study few-body systems in atomic, nuclear and molecular physics \cite{mitroy2013theory, suzuki1998stochastic}. The wave function is expanded in terms of Gaussian basis functions as
\begin{align}
\ket{\psi}=\sum_{i=1}^{K}c_i \hat{\mathcal P}\ket{g_i},
\end{align}
where $c_i$ is a linear variational parameter, $\ket{g_i}$ is the Gaussian, $\hat{\mathcal P}$ is a symmetry operator imposing the proper symmetries and $K$ is the number of Gaussians. Inserting this into the Schr\"odinger equation, $\hat{H}\ket{\psi}=E\ket{\psi}$, and subsequently multiplying from the left by $\bra{g_j}$ leads to the generalized eigenvalue problem
\begin{align}\label{eq:gennv}
\mathcal H \bm c = E \mathcal N \bm c,
\end{align}
where $\bm c=(c_1, c_2,...,c_K)^\text{T}$, $\mathcal H_{ij}=\bra{g_i}\hat{H}\hat{\mathcal P}\ket{g_j}$ and $\mathcal N_{ij}=\bra{g_i}\hat{\mathcal P}\ket{g_j}$. In order to simplify the matrix elements we have used that $[\hat H,\hat{\mathcal P}]=0$ and $\hat{\mathcal P}^2=\hat{\mathcal P}$. We define $\hat{\mathcal P}=\hat\Pi \hat{\mathcal{S}}$, where $\hat \Pi$ imposes the desired parity and $\hat{\mathcal S}$ is a symmetrizer or antisymmetrizer ensuring the correct quantum statistics under exchange of the majority particles. Equation \eqref{eq:gennv} is easily solved performing a Cholesky decomposition \cite{Golub1996Matrix} on $\mathcal N$ and subsequently rewriting it to an ordinary eigenvalue problem. This returns the linear variational parameters along with the energy spectrum.

As we consider systems that are highly anisotropic the basis functions, $\ket{g}$, have to include a directional dependence. We, therefore, choose to employ the so-called fully correlated Gaussians which takes the form
\begin{align}\label{eq:gauss_basis}
\braket{\bm x}{g}=e^{-\bm x^\text{T} A \bm x+\bm s^\text{T}\bm x}
\end{align}
where $A$ is a positive definite correlation matrix of size $3(N-1)\cross 3(N-1)$ and $\bm s$ is a shift vector of size $3(N-1)$, both holding non-linear variational parameters. The basis is build using stochastic optimization after which a Nelder-Mead algorithm \cite{Teukolsky2007Numerical} is used to perform multiple refinement cycles.

Expanding the wave function in terms of Gaussians has the advantage that all matrix elements are analytically expressible, enabling for extensive numerical optimization and therefore accurate results. The necessary matrix elements are found in \appref{app:matrix-elements}. A disadvantage is the non-orthogonality of the basis functions. To ensure numerically accurate results it is necessary to require linear independent basis functions. See \cite{mitroy2013theory,suzuki1998stochastic} for a thorough introduction to the correlated Gaussian method.

\section{1+1 System} \label{sec:1+1}
To ensure that the CGM implementation is done properly we test it on a system with known solutions. Here we chose the 1+1 systems, i.e. two distinguishable particles interacting in a quasi-1D harmonic trap. The 1+1 system has been solved exactly in both 1D \cite{busch1998two, farrell2009universality} and in an axial symmetric trap \cite{idziaszek2005, idziaszek2006analytical}.

Before we can present the results we have to specify the units by choosing a mass scale $\mu$. Here we use the usual two-body reduced mass $\mu=\tfrac{m_Mm_I}{m_M+m_I}$, where $m_M$ and $m_I$ denote the mass of the majority and impurity particles respectively. These units will also be used for the 2+1 system.
\subsection{Energies}
\begin{figure}[ht]
	\centering
	\includegraphics[]{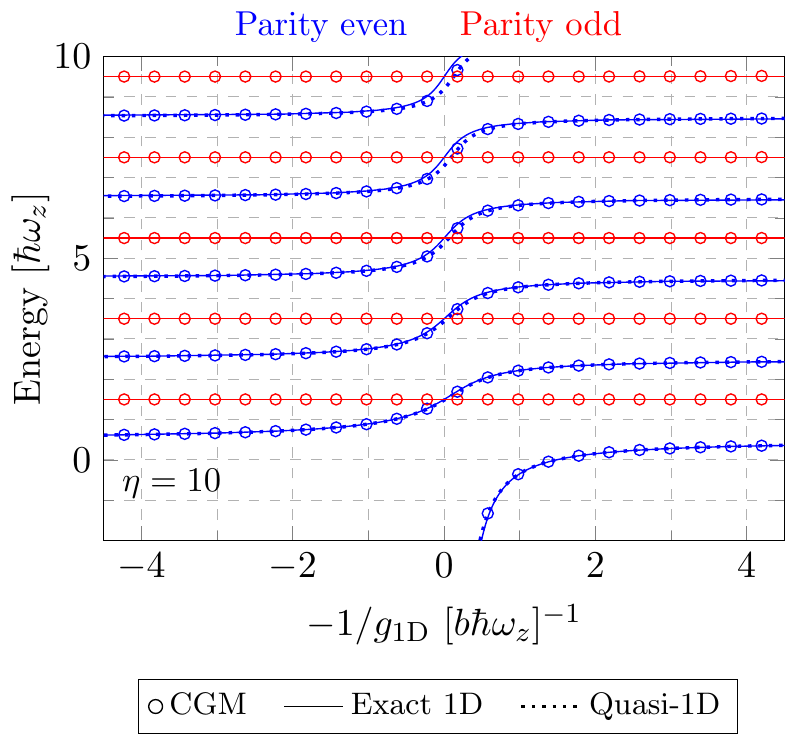}
	\caption{\label{fig:1+1_Energy}Energy spectrum for the 1+1 system as a function of the one-dimensional coupling constant $g_\text{1D}$, excluding the center-of-mass contribution. The solid and dashed lines are exact solutions to the 1D and quasi-1D system respectively. The circles are numerically calculated using CGM. For clarity the ground state energy of the transverse trap, $E_\perp=\eta$, has been subtracted from the energy obtained solving the relative Hamiltonian.}
\end{figure}

The energy spectrum for the relative motion, as a function of $1/g_\text{1D}$, is shown in \figref{fig:1+1_Energy}. 
The solid lines are the exact solution to the 1D system while dashed lines are the exact quasi-1D solution found using $\eta=10$. We have subtracted the energy of the transverse ground state for clarity. It is seen that low lying states have nearly exact 1D behaviour for an aspect ratio of $\eta=10$ while a clear deviation appears for excited states around the fermionization point, i.e. $1/g_\text{1D}=0$. Deviation from the 1D solution is expected for excited states as these are closer to the transverse excitation energy, $E_\perp=\eta$.
 
The circles in \figref{fig:1+1_Energy} are obtained using CGM and show excellent agreement with the exact quasi-1D solution even for highly excited states. Notice how odd parity states are not influenced noticeably by the changing interaction. This is because the relative wave function vanishes as the interparticle distance approaches zero. As the interaction is short-range the particles only feel each other very little and the energy therefore stays virtually constant. For the exact solutions which used zero-range interactions the particles will not feel each other at all. On the other hand, the even parity states interact and are therefore of particular interest.

\subsection{Axial Density}
\begin{figure}[ht]
	\centering
	\includegraphics[]{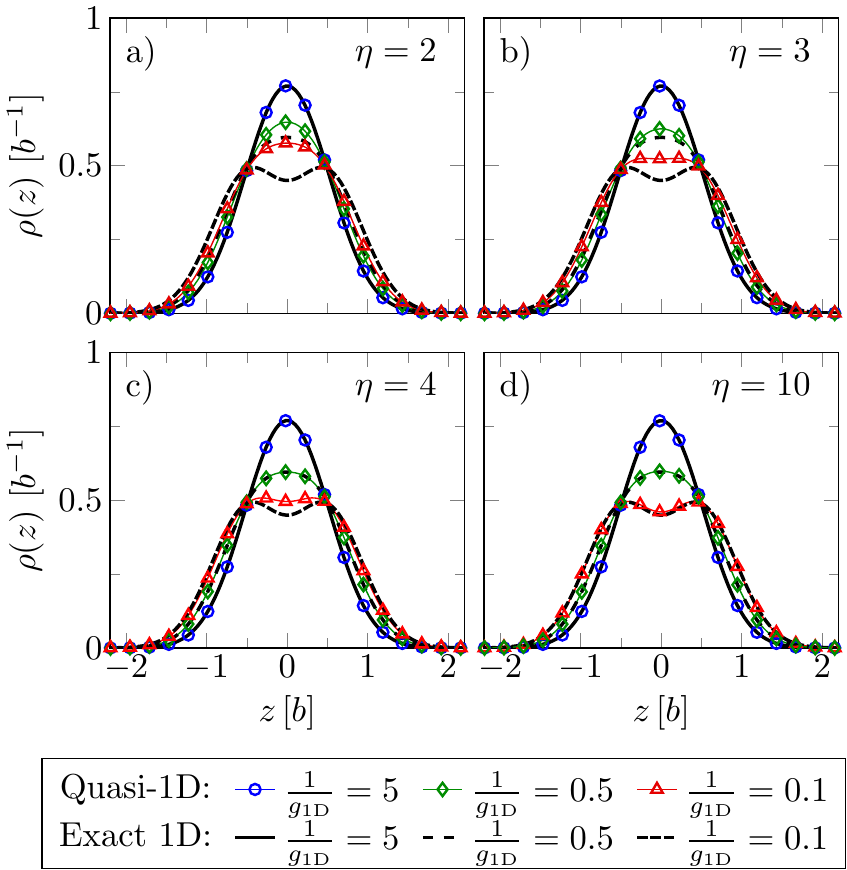}
	\caption{\label{fig:1+1_one_body_density}Ground state density distribution in the z-axis, for the mass-balanced 1+1 system. Panels (a)-(d) correspond to different aspect ratios (see figure). The black lines are exact 1D solutions while the circles are found using CGM. The densities are normalized to an area of one.}
\end{figure} 
The 1D behaviour of the system can be studied using the density distributions in the $z$-axis and compared to exact 1D results. We will only consider the even parity ground state which moves continuously from the repulsive to the attractive side (see \figref{fig:1+1_Energy} for clarity). This can be seen in \figref{fig:1+1_one_body_density}, where solid, dotted and dashed lines are exact 1D densities while marked lines are numerically calculated. The definition of the density distribution can be seen in \appref{app:density-distribution}. Panel (a) shows the density for a small anisotropy, i.e. $\eta=2$, where only the weak interaction, $1/g_\text{1D}=5$, agrees with the exact 1D solution. This show that for $1/g_\text{1D}=0.5$ and $1/g_\text{1D}=0.1$, interacting systems can use the transverse dimensions and does not behave 1D. For $\eta=4$, seen in panel (c), the density distribution for $1/g_\text{1D}=0.5$, surprisingly follow the exact 1D closely. This indicates that such systems experience 1D properties already at small aspect ratios, even for relatively strong interactions. Moving on to $\eta=10$, the densities now follow the exact 1D closely with only a small deviation for $1/g_\text{1D}=0.1$ indicating that the system is effectively 1D.

\subsection{Occupation Numbers}
\begin{figure}[ht]
	\centering 
	\includegraphics{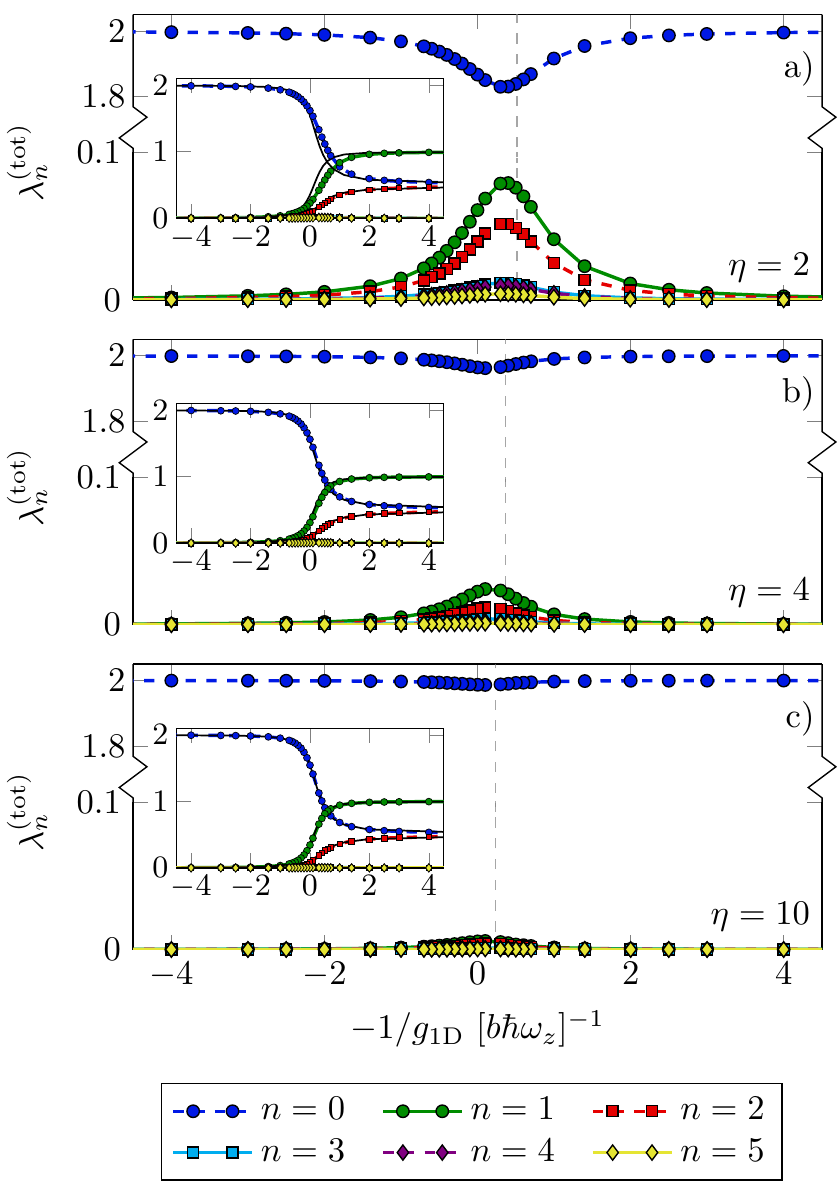}
	\caption{\label{fig:1+1_occupation}Total occupation number, $\lambda^{(tot)}_{n}=\lambda^{(M)}_{n}+\lambda^{(I)}_{n}$, in the transverse direction in terms of $1/g_\text{1D}$ for the ground state moving continuously from the repulsive to the attractive side (See \figref{fig:1+1_Energy}). The different panels corresponds to different aspect ratios as seen in the figure. The insert show the corresponding occupation number in the $z$-axis. Data points are connected with lines for clarity. The solid black line in the insert show the exact 1D occupation numbers calculated from the analytical wave function.}
\end{figure}
A new approach to study the crossover from 3D to 1D will now be formulated and discussed. In particular, the occupation numbers of the transverse states will be plotted for different values of $\eta$. These are introduced in \appref{app:One-Body-Density-Matrix} as the eigenvalues, $\lambda$, of the one-body density matrix in \equref{eq:one-body-density-matrix}. In \figref{fig:1+1_occupation} we plot the total occupation, $\lambda^{(tot)}_{n}=\lambda^{(M)}_{n}+\lambda^{(I)}_{n}$, of the first six single particle states, i.e. $n\in \{0,1,...,5\}$, as a function of $1/g_\text{1D}$. We do this for different aspect ratios as seen in panels (a), (b) and (c). Notice the discontinuous $y$-axis, introduced for clarity. The inserts show the corresponding occupation in the $z$-direction. We observe that for $\abs{1/g_\text{1D}}\gtrsim2.5$, i.e. weak interactions, the particles are located in the lowest single particle state. As the interaction strength increases a clear peak is seen for excited transverse states near $1/g_\text{1D}=0$ for $\eta=2$. This show that the system is no longer limited to only the single-particle ground state. For exact 1D systems, only one such state will be present. For $\eta=4$ this effect decreases significantly and nearly vanishes for $\eta=10$, indicating that the system is effectively 1D. Notice also that the occupation number does not peak at $g_\text{1D}=\infty$ nor at $a_\text{3D}=\infty$ which is shown as the vertical dashed line. In terms of $S_{ij}$, $g_\text{1D}=\infty$ is of no particular interest and it is therefore not surprising that the peak does not occur at this point. It should be noted that even for low anisotropies and strong interaction, the occupation of excited transverse states is small.

Notice also that the occupation in z-direction does not change significantly for the three aspect ratios, nevertheless, as $1/g_\text{1D}$ changes continuously from positive to negative, excited single-particle states become occupied. The solid black line in the insert shows the exact 1D occupation numbers calculated from the analytical wave function. A small deviation is seen for $\eta=2$ near $1/g_\text{1D}=0$. \\\hspace*{\fill}

Before continuing to the 2+1 system a few things should be noted. First, the CGM implementation can describe the system to a very high degree of accuracy. Second, correspondence between numerically obtained results and the idealized 1D model is highly dependent on the interaction strength and the aspect ratio. For $g_\text{1D}<2$ the 1+1 the density distributions are indistinguishable from the exact 1D for aspect ratios as low as $\eta=4$. These results will be compared to those presented in the next section, which will give information about the dependence on particle number. 

\section{2+1 System}\label{sec:2+1} 
In this section we study the properties of the 2+1 system in the dimensional crossover from 3D to 1D. We start by considering the radial distribution for different aspect ratios and interaction strength, next we discuss the axial densities and at last the occupation numbers. In \appref{app:strict-1D} we thoroughly study this system in the 1D limit where $\eta=50$ and compare to the idealized model.
\subsection{Radial Distribution}
\begin{figure}[htbp]
	\centering
	\includegraphics{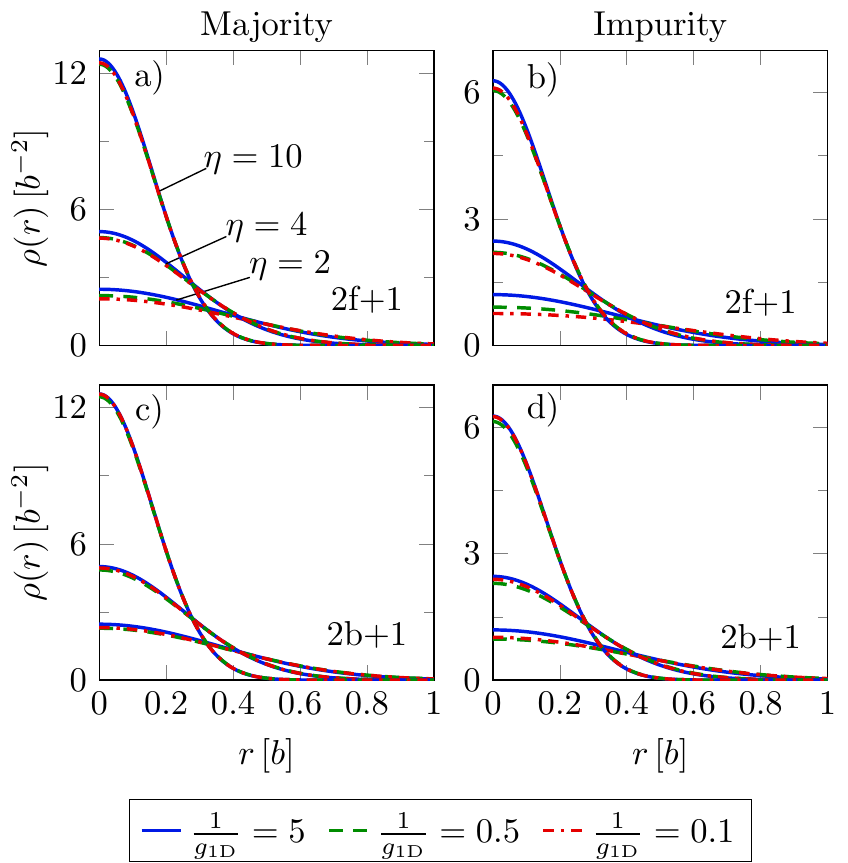}
	\caption{\label{fig:Radial_dist_eta}Radial density distribution in term of $r=\sqrt{x^2+y^2}$, for different aspect ratios and interaction. The first row corresponds to the 2f+1 system and the second to the 2b+1, while the columns to the majority and impurity respectively. As the order of the distributions does not change with respect to the aspect ratio, these are only labeled in panel (a).}
\end{figure}

The radial distributions for the 2f+1 and 2b+1 system are shown in \figref{fig:Radial_dist_eta}. These are calculated using \equref{eq:general-form-factor-element} and \appref{app:density-distribution} to obtain information about the density in the transverse directions, i.e. the $xy$-plane. These were changed to polar coordinates, i.e. $r=\sqrt{x^2+y^2}$ and $\theta$, and the angular integration was performed (see xy-plane in \figref{fig:2+1_system}). It is normalized according to $N_{c}=2\pi \int \rho_{c}(r) r \dd{r}$, where $N_{c}$ is the number of particles and $\rho_{c}(r)$ is the radial density for $c=\text{I},\text{M}$. A general feature for all panels is that, as the aspect ratio is increased, the difference between the weakly and strongly interacting densities decreases. This shows that the transverse dimensions are of less importance for higher aspect ratios. 

For the 2f+1 system, it is observed that large interaction strength leads to more radial delocalization, especially for $\eta=2$. In particular the impurity shows a large delocalization for $1/g_\text{1D}=0.1$ compared with $1/g_\text{1D}=5$. The same is seen for the 2b+1 system although to a smaller degree. Here the densities depend less on the interaction strength indicating that a bosonic system will experience 1D behaviour at a lower aspect ratio than a fermionic system. Intuitively this is clear as the fermionic system is higher in energy, i.e. closer to the transverse excitation energy. Also, the strongly interacting case for the 2b+1 system is more localized radially compared to $1/g_\text{1D}=0.5$ density, albeit by a small amount.

\subsection{Axial Density}
\begin{figure}[t]
	\centering
	\includegraphics[]{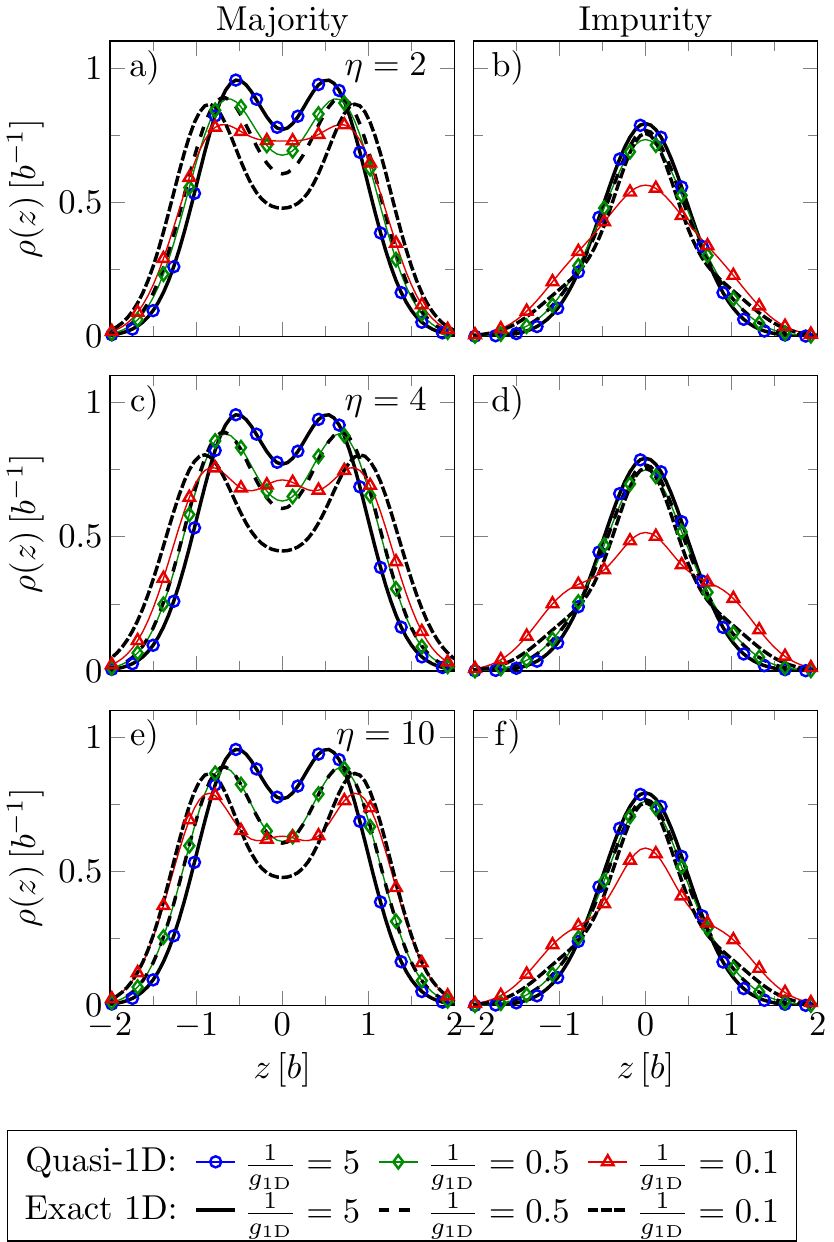}
	\caption{\label{fig:density_2f+1_z}2f+1 ground state density distributions for different interaction strength and aspect ratios. Solid lines are density distribution using the idealized model, i.e. 1D and zero-range interactions. The three rows correspond to the aspect ratio seen in panels (a), (c) and (e). }
\end{figure}

\begin{figure}[t]
	\centering
	\includegraphics[]{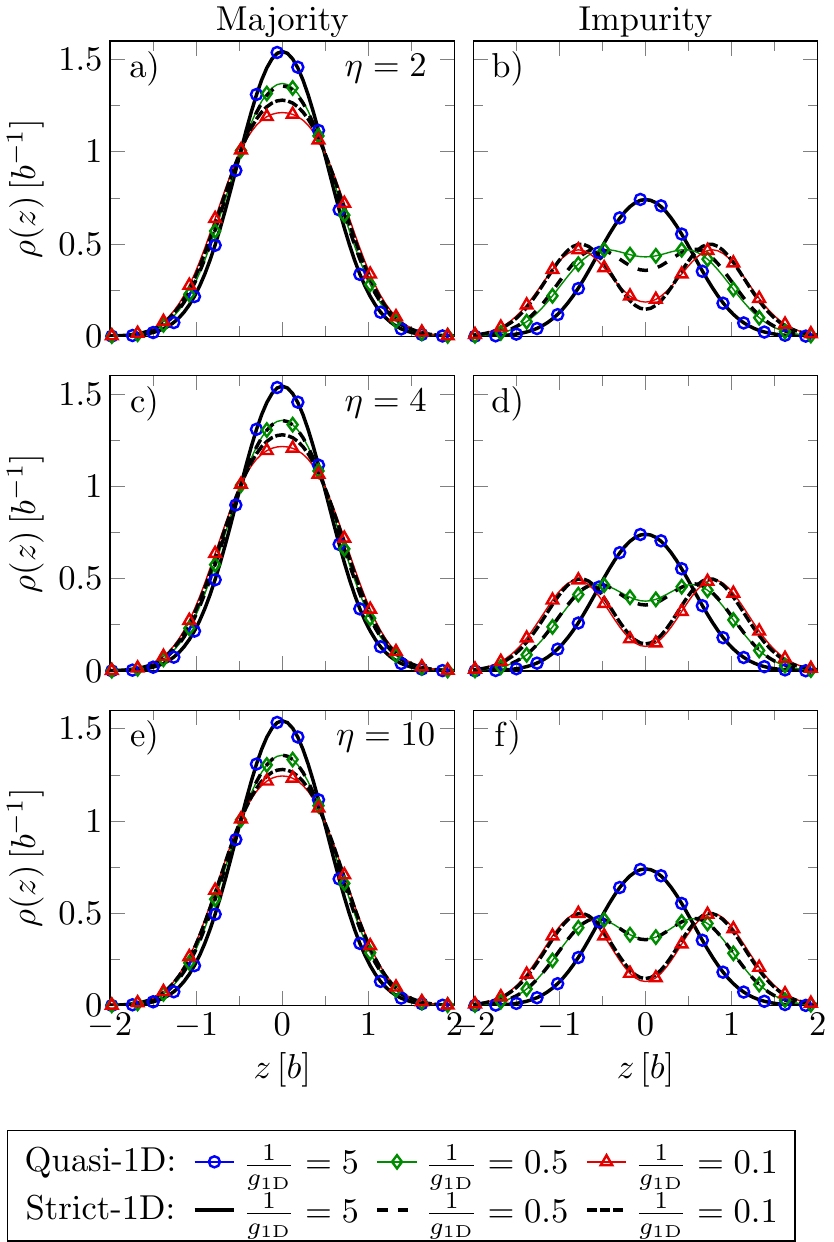}
	\caption{\label{fig:density_2b+1_z}The same as \figref{fig:density_2b+1_z} but for the 2b+1 system.}
\end{figure}

The same features are seen in the one-body densities, along the z-axis, in \figref{fig:density_2f+1_z} and \figref{fig:density_2b+1_z}.
Specifically, \figref{fig:density_2f+1_z} show the 2f+1 one-body density. In panels (a) and (b) it is seen that for weak interactions, i.e. $1/g_\text{1D}=5$, the density distribution agrees with the 1D case. Increasing the interaction to $1/g_\text{1D}=0.5$ a clear deviation is present compared to the 1D density, indicating that the system no longer behaves like a 1D system. This deviation is further increased for $1/g_\text{1D}=0.1$. From \appref{app:strict-1D} it is known that the 2f+1 ground state has the particle ordering $\uparrow\downarrow\uparrow$. Using this knowledge, it is intuitively clear why the majority density increases toward $z=0$, compared to the 1D case, as the majority particles can move in the radial trap appearing closer in the z-direction. For $\eta=4$ the same features are seen nevertheless for $1/g_\text{1D}=0.5$ the density is virtually indistinguishable from the 1D case for both the majority and impurity. Interestingly this shows that, unless the system is strongly interacting, the 2f+1 ground state behaves 1D for aspect ratios as low as $\eta=4$. Remember \figref{fig:1+1_one_body_density} where the same conclusion was made for the 1+1 system, yet here the deviation for $1/g_\text{1D}=0.1$ vanished already for $\eta=10$. As seen in panels (e) and (f), where the densities are shown for $\eta=10$, this is not the case for the 2f+1 system.

A different behaviour is seen for the 2b+1 system. For $\eta=2$ the majority density follows closely the 1D case, while the impurity shows a small deviation. Compared to the 2f+1 system, this shows that a bosonic system experiences 1D properties for lower aspect ratios than fermionic. It is interesting to note that for $\eta=4$ the $1/g_\text{1D}=0.1$ density for the impurity is indistinguishable from the exact 1D while a small deviation is seen for $1/g_\text{1D}=0.5$. This behaviour was also noticed in the radial densities seen in \figref{fig:Radial_dist_eta}. For $\eta=10$ the impurity densities experience 1D behaviour while a small deviation is seen for the majority particles.

In general \figref{fig:density_2f+1_z} and \figref{fig:density_2b+1_z} show that the 2+1 system can be assumed 1D for aspect ratios as low as $\eta\sim 4$ if $1/g_\text{1D}>0.5$.

\subsection{Occupation Numbers}
\begin{figure}[t]
	\centering
	\includegraphics[]{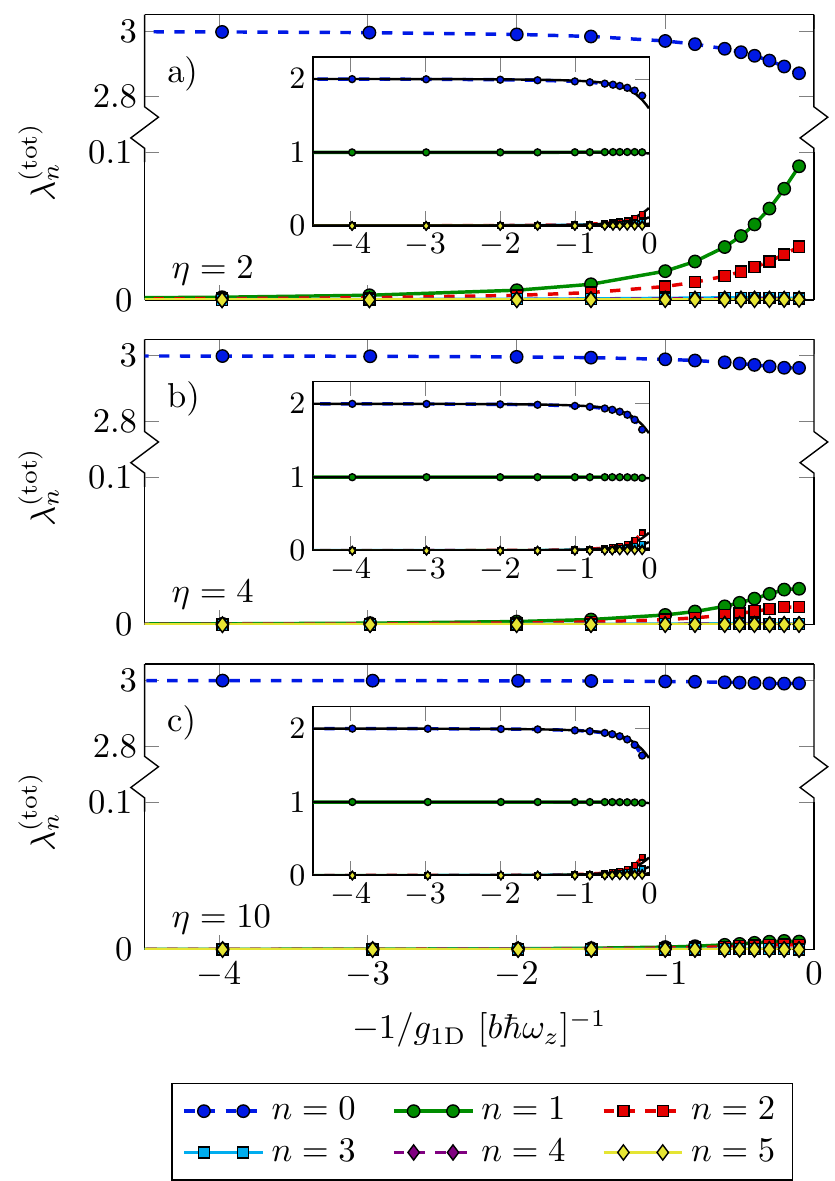}
	\caption{\label{fig:occupation_state_1_fermion_odd}Total occupation number, $\lambda^{(\text{tot})}_n=\lambda^{(\uparrow)}_n+\lambda^{(\downarrow)}_n$, as a function of the interaction strength for the 2f+1 ground state. It shows the first six single particle states obtained by diagonalization of the one-body density matrix in \equref{eq:one-body-density-matrix}. Data points are connected with lines for clarity. Notice the discontinuous $y$-axis in all panels. The insert shows the $z$-axis occupation numbers where the solid black line corresponds to exact 1D.}
\end{figure}

\begin{figure}[t]
	\centering
	\includegraphics[]{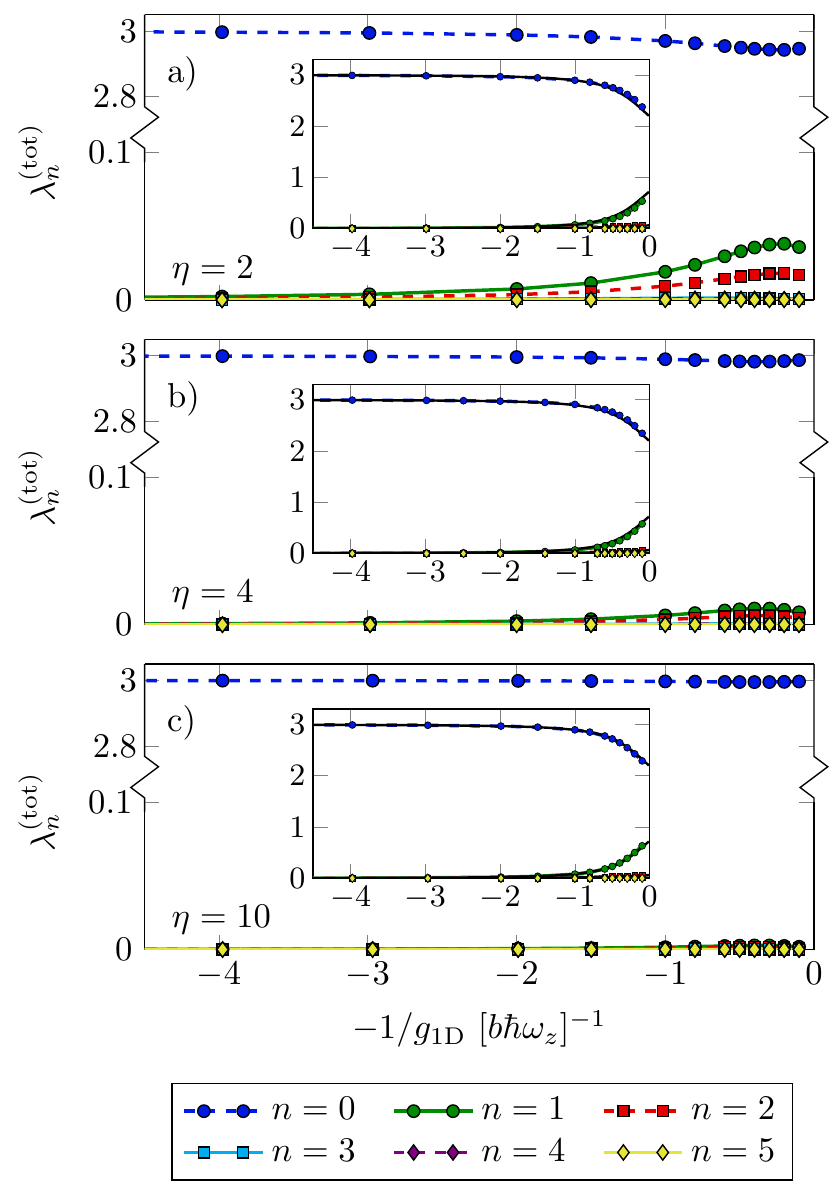}
	\caption{Shows the same as \figref{fig:occupation_state_1_fermion_odd} but for the 2b+1 system i.e. $\lambda^{(\text{tot})}_n=\lambda^{(A)}_n+\lambda^{(B)}_n$.}
	\label{fig:occupation_state_1_boson_even}
\end{figure}

Apart from density distributions, the crossover from 3D to 1D can be studied through the one-body density matrix as shown for the 1+1 system. \figref{fig:occupation_state_1_fermion_odd} shows the 2f+1 ground state occupation in the transverse directions for a range of interactions.
The insert is the corresponding occupation in the z-direction. The three rows correspond to $\eta=2,\, 4$ and $10$, respectively. Due to numerical limitations, it is not possible to study the attractive side. Interestingly a clear occupation of the first and second transverse excited states is seen for $\eta=2$ near $1/g_\text{1D}=0$. This shows that strongly interacting particles can move in the transverse direction, albeit by a small amount. Increasing the aspect ratio to $\eta=4$ the occupation decreases and nearly vanishes for $\eta=10$. The insert shows the occupation in the z-axis which does not change significantly for the three aspect ratios. Also, the fermionic nature of the majority particles is seen in the inserts as two single-particle states are occupied.
The 2b+1 system seen in \figref{fig:occupation_state_1_boson_even} shows a different behaviour. Here the occupation in the transverses direction seems to have peaked on the repulsive side. Furthermore, it is seen that the bosonic system does not occupy excited transverse states as much as the fermionic system. As for the 2f+1 system, the occupation in the $z$-direction does not change significantly for the different aspect ratios. 

\section{Conclusion and Outlook}\label{sec:Con_and_out}
We have studied 1D systems using a full 3D treatment to analyze the dimensional crossover from 3D to 1D and the validity of the idealized 1D model. We used the 1+1 system as a benchmark for the CGM implementation and showed that for $\eta=10$ the quasi-1D system agrees with the idealized model for all interactions. Further we showed that the occupation numbers in the transverse direction can be used to study the dimensionality of the system. In the following section we conclude on the observation for the 2f+1 and 2b+1 systems along with an outlook presenting further studies.
\subsection{Conclusion}
We studied both the 2f+1 and 2b+1 systems and showed that the 2b+1 follows the idealized 1D model closely for $\eta\geq4$. Surprisingly this indicates that the bosonic system experiences 1D behaviour for low anisotropy even in the strongly interacting limit. On the other hand, the fermionic system showed large deviations from the 1D densities for $\eta=10$ and $1/g_\text{1D}=0.1$. This deviation is even visible for $\eta=50$ as shown in \appref{app:strict-1D}. The weaker interacting systems were in agreement with the 1D model. The same tendency was observed in the transverse occupation numbers. Here the fermionic system occupied excited transverse states significantly more than the bosonic system. From this, it is seen that the fermionic systems need a large aspect ratio to reproduce the 1D results compared with the 1+1 and 2b+1 system. Also, even though only two and three-particle systems are considered, the results indicate that the particle number is of great importance for fermionic systems.

An interesting feature was also seen in the occupation numbers for the 2b+1 system compared with the 1+1 and 2f+1 systems. The 1+1 has a peak in the transverse occupation of excited states on the attractive side of the spectrum. The 2f+1 system seems to have the same behaviour while the 2b+1 system has a peak on the repulsive side.

In summary, it was observed that systems with $1/g_\text{1D}>0.5$ could be considered approximately 1D for $\eta\geq4$ irrespectively of the particle types or system size. For strong interaction, $1/g_\text{1D}=0.1$, the 1+1 and 2b+1 systems showed 1D behavior at $\eta=10$ in contrast to the 2f+1 system which showed a clear deviation from the 1D case.

\subsection{Outlook}
The implemented version of the correlated Gaussian method can, in principle, quickly and easily be applied to other systems with more particles, different trap configurations, different interactions and so on. Further studies which can be done with the current implementation include excited states and mass-imbalanced systems. As excited states are higher in energy and, in general, have a larger spatial extent these are expected to depend more on the trap configuration compared to the ground state. This may result in interesting phenomena which could be investigated.

A natural continuation of our work would be to add a particle more and redo the analysis. In particular, the 2f+1 system showed interesting properties in the strongly interacting limit. Studying the 3f+1 system would, therefore, be of interest as we expect this to depend more on the trap configuration.

Another venue of interest is to compare the present techniques to the results obtained from interpolatory ansatz wave functions for few-body systems. This is discussed for both static \cite{andersen2016interpolatory, pkecak2017four, lindgren2019systematic} and dynamic \cite{kahan2019driving} cases.

\appendix

\section{Matrix elements}\label{app:matrix-elements}
An advantage of the correlated Gaussian method is that all matrix elements can be calculated analytically and that the complexity does not increase for increasing particle number \cite{Fedorov2016}. This appendix will give the necessary matrix elements to perform the calculation in the main text. 

Consider a system of $N$ particles in $D$ dimensions. We solve the Schr\"odinger equation, for the relative motion, by expanding the wave function as a linear combination of Gaussians, $\braket{\bm x}{g}=\exp(-\bm x^\text{T}A\text{T}+\bm s^\text{T}\bm x)$. Here $A$ and $\bm s$ contains variational parameters and are of size the $D(N-1)\cross D(N-1)$ and $D(N-1)$ respectively. The first matrix element is the overlap between the Gaussian $\ket{g'}$ with parameters $A'$, $\bm s'$ and $\ket{g}$ with parameters $A$, $\bm s$ given as
\begin{align}\label{eq:overlap}
	\braket{g'}{g}=e^{\frac{1}{4}\bm v^\text{T} B^{-1} \bm v} \frac{\pi^{\frac{D(N-1)}{2}}}{\sqrt{\det(B)}}\equiv M,
\end{align}
where $\bm v=\bm s'+\bm s$ and $B=A'+A$. Here $N-1$ is the reduced number of particles since Jacobi coordinates are used. Due to the Gaussian nature, the necessary matrix elements can be calculated from \equref{eq:overlap}. The kinetic matrix element is
\begin{align}
\begin{split}
\bra{g'}-\grad_{\bm x}^\text{T} \Lambda \grad_{\bm x}\ket{g}=&M\Big(2 \tr(A'\Lambda A B^{-1})\\
&+(\bm s'-2A'\bm u)^\text{T}\Lambda (\bm s-2A \bm u)\Big),
\end{split}
\end{align}
where $\bm u=\tfrac{1}{2}B^{-1}\bm v$. The calculation of the kinetic term trivially leads to the oscillator matrix elements on the form
\begin{align}
	\bra{g'}\bm x^\text{T} \Omega \bm x \ket{g}=\left(\bm u^\text{T} \Omega \bm u+\frac{1}{2} \tr(\Omega B^{-1})\right)M.
\end{align} 
The last matrix element is the Gaussian interaction. It takes the form
\begin{align}
	\bra{g'}e^{-\gamma \bm x^\text{T} W_{ij} \bm x}\ket{g}=e^{\frac{1}{4} \bm v^\text{T}B'^{-1}\bm v} \frac{\pi^{\frac{D(N-1)}{2}}}{\sqrt{\det(B')}},
\end{align} 
where $B'=B+\gamma W_{ij}$ and 
\begin{align}
	W_{ij}=\sum_{c=x,y,z} w_{ij,c} w_{ij,c}^\text{T}.
\end{align} 
 Here $w_{ij,c}$ is the vector fulfilling 
\begin{align}
	r_{i,c}-r_{j,c}=w_{ij,c}^\text{T}\bm x,
\end{align} 
which is easily generated using \equref{eq:Jacobi-Matrix}.

\subsection{General Form-Factor}\label{app:general-form-factor}
The following calculation is carried out using particle coordinates, $\bm r$, yet it generalizes trivially to Jacobi coordinates using \equref{eq:Jacobi-Matrix}, \equref{eq:CM-solution} and $\Psi(\tilde{\bm x})=\psi(\bm x) \phi(\vec{x}_N)$  
For a general potential $V(r_i)=V(w_i^\text{T}\bm r)$ with Fourier transform 
\begin{align}
V(w_i^\text{T}\bm r)=\int \frac{\dd{k}}{2\pi} f(k)e^{ikw_i^\text{T}\bm r},
\end{align}
the matrix element reduces to the evaluation of a one-dimensional integral. Given an index set $S\subseteq \{1,...,DN\}$ of particle coordinates this generalizes to
\begin{align}\label{eq:Matrix_element_of_product_of_potentials}
\begin{split}
\bar{M}&=\bra{g'}\prod_{i\in S} V(w_i^\text{T}\bm r)\ket{g}\\
&=\left(\prod_{i\in S} \int \frac{\dd k_i}{2\pi} f(k_i)\right)\hspace{-0.1cm} \int \dd{\bm r}e^{-\bm r^\text{T} B \bm r+\left(\bm v+ \sum_{j\in S} ik_j w_j\right)^\text{T} \bm r}.
\end{split}
\end{align}
The spatial integral is equivalent to the overlap from \equref{eq:overlap} leading to
\begin{align}\label{ME-general-potential}
\begin{split}
\bar{M}=&\left(\prod_{i\in S} \int \frac{\dd k_i}{2\pi} f(k_i)\right)\cross\\
& e^{\frac{1}{4} \left(\bm v +\sum_{j\in S} i k_j w_j\right)^\text{T}B^{-1}\left(\bm v +\sum_{j\in S} i k_j w_j\right)}\frac{\pi^{\frac{DN}{2}}}{\sqrt{\det(B)}},
\end{split}
\end{align} 
which, using the definitions
\begin{align}
C_{ij}\equiv\frac{1}{4} w_i^\text{T} B^{-1} w_j \quad \quad \text{and} \quad \quad q_i\equiv\frac{1}{2}w_i^\text{T} B^{-1}\bm v,
\end{align}
becomes
\begin{align}
\bar{M}=M \left(\prod_{i\in S} \int \frac{\dd k_i}{2\pi} f(k_i)\right) e^{-\bm k^\text{T} C \bm k+i\bm q^\text{T} \bm k }.
\end{align}
Here $\bm k$ is a vector of size-$\abs{S}$ holding the integration variables, $\bm q=\{q_i\}$ is a size-$\abs{S}$ column and $C=\{C_{ij}\}$ a $\abs{S}\cross\abs{S}$ symmetric positive-definite matrix. This integral is now expressible in terms of the potential using $\abs{S}$ inverse Fourier transforms as
\begin{align}\label{eq:general-form-factor-element} 
\bar{M}=&M \left(\prod_{i\in S} \int \frac{\dd k_i}{2\pi} \int \dd r_i V(r_i) \right) e^{-\bm k^\text{T} C \bm k+i\left(\bm q- \bm r_S\right)^\text{T} \bm k }\nonumber\\
\begin{split}
=&\frac{M}{(2\pi)^{\abs{S}}} \frac{\pi^\frac{\abs{S}}{2}}{\sqrt{\det(C)}} \left(\prod_{i\in S} \int \dd{r_i} V(r_i)\right)\cross\\
&e^{-\frac{1}{4}\left(\bm q-\bm r_S\right)^\text{T} C^{-1}\left(\bm q-\bm r_S\right)} 
\end{split}
\end{align}
where $\bm r_S=\{r_i\}$. This expression is particular suited for the calculation of density functions which will be shown in the following section.

\section{Density Distributions}
In order to study how the particle are distributed in the trap we define the density distributions. These are defined as the usual probability amplitude yet, at we consider multidimensional wave function, coordinates of no interest are integrated out.  
\subsection{n-Body Density}\label{app:density-distribution}
We define the one-body density distribution as the integral
\begin{align}\label{eq:one-body-density}
	\rho_i(r)=\int \dd{\bm r}\delta(r_i-r) \abs{\Psi(\bm r)}^2,
\end{align}
i.e. the integration is performed over all coordinates except the $i$'th. This can be used to obtain information about the majority and impurity distribution in the trap individually. The integral in \equref{eq:one-body-density} is easily calculated using \equref{eq:general-form-factor-element}. To obtain more information about the system we define the so-called pair correlation function as
\begin{align}\label{eq:pair-correlation}
	\rho_{ij}(r,r')=\int \dd{\bm r} \delta(r-r_i)\delta(r'-r_j) \abs{\Psi(\bm r)}^2,
\end{align}
which can be used to obtain information about the correlation between the majority particles and the impurity. More generally an n-body density distribution can be calculated using a general density operator on the form
\begin{align}
	\mathcal O_n=\prod_{i\in S} \delta(r'_i-w_i^\text{T}\bm r),
\end{align} 
where $n=\abs{S}$ and $S\subseteq \{1,...,3N\}$ is and index set. Inserting this into \equref{eq:general-form-factor-element} the $n$-body density follows directly.
\subsection{One-Body Density Matrix}\label{app:One-Body-Density-Matrix}
The one-body density matrix is a generalization of the one-body density. It is defined as
\begin{align}
\scriptstyle
\begin{split}\label{eq:one-body-density-matrix}
	\rho(r_q,r'_q)=&\int \Psi(r_1,..., r_q,..., r_{3N})\Psi^\dagger(r'_1,..., r'_q,..., r'_{3N})\\
	&\prod_{i\neq q}^{3N}\dd{r_i}\prod_{j\neq q}^{3N}\dd{r'_j},
\end{split}
\end{align}
where $r_i$ for $i \in \{1,2,3\}$ is the elements in the position vector for the first particle,  $i\in \{4,5,6\}$ for the second and so on. The one-body density matrix can be calculated analytically using the operator
\begin{align}\label{OBDM}
\mathcal{O}_{rr'}^{(i)}=\delta(w_i^\text{T}\bm  r-r)\delta(w_i^\text{T}\bm r'-r')\prod^N_{j\neq i} \delta(w_j^\text{T}\bm r-w^\text{T}_j \bm r'),
\end{align}
as
\begin{align}
	\rho(r,r')_i=\int \dd{\bm r}\dd{\bm r'}\mathcal{O}^{(i)}_{rr'}\Psi(\bm r)\Psi^\dagger(\bm r')
\end{align}
following the same procedure as in \appref{app:general-form-factor}. It can be diagonalized,  
\begin{align}
\int \rho(r,r') \chi_n(r') \dd{r'}=\lambda_n \chi_n(r), 
\end{align}
to obtain the single-particle states $\chi_n(r)$ and eigenvalues $\lambda_n$. The eigenvector and eigenvalues are known as the natural orbitals and occupations numbers respectively. As the density matrix is both hermitian and positive definite the occupation numbers are real and positive. For non-interacting systems in a Harmonic potential, $\chi_n(r)$ is the eigenfunctions and $\lambda_n$ is the corresponding occupation \cite{pethick2002bose}. The eigenvalues are then interpreted as the occupation of the single particles states. The normalization is often chosen such that $\sum_n \lambda_n^{(I/M)}=N_{I/M}$, i.e. the sum of occupation numbers equals the number of particles in the subsystem. We define the total occupation as $\lambda^{(tot)}_{n}=\lambda^{(M)}_{n}+\lambda^{(I)}_{n}$.

The natural orbitals and occupation numbers are easily found by discretization and subsequent diagonalization of $\rho(r,r')$, introducing a remarkable opportunity to study the single particles states, and their occupation, in the transverse and axial directions. This can be used to determine the effect of the transverse trap in a quasi-1D system, as in true 1D only one state should be occupied.

\section{2+1 in 1D limit}\label{app:strict-1D}
\begin{figure}[htp]
	\centering
	\begin{tikzpicture}
	\node[] at (-6.6,3.5){a)};
	\node[] at (-1.5,3.5){b)};
	\node[inner sep=0pt] (whitehead) at (-3.5,0)
	{\includegraphics[]{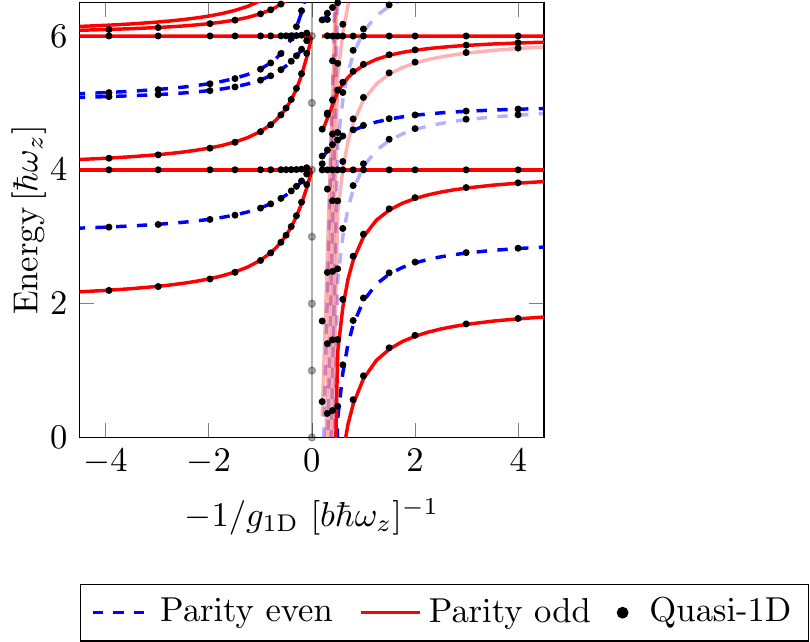}};
	\node[inner sep=0pt](whitehead) at (-0.6,0.42){\includegraphics[]{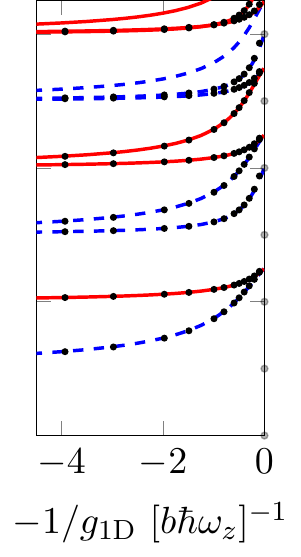}};
	\end{tikzpicture}
	\caption{\label{fig:2f+1_massbalanced_spectrum}Panel (a) shows the intrinsic energy for the 2f+1 system. Lines show the energy of the idealized 1D model with zero range interaction, where bound states are dimmed for clarity. Marks are quasi-1D where the transverse ground state energy, $E_\perp=2\eta$, has been subtracted to allow for direct comparison between the two models. Panel (b) shows the same as panel (a) but for the 2b+1 system and only for repulsive interactions.}
\end{figure} 
In this section we study the 2f+1 and 2b+1 systems thoroughly in the 1D limit i.e. $\eta=50$ and compare to the idealized 1D model with zero range interactions.

The energy spectrum is seen in \figref{fig:2f+1_massbalanced_spectrum}. Here lines correspond to the idealized 1D model while marks are quasi-1D. It shows the total intrinsic energy i.e. the total energy minus $\left(\eta+1/2\right)\hbar\omega_z$, from the center-of-mass. Further, the transverse ground state energy, $E_\perp=2\eta$, is subtracted from the quasi-1D solution to allow for direct comparison between the two models. Panel (a) shows the first nine states of the 2f+1 system where, due to numerical limitations, interactions larger than $g_\text{1D}\sim 10$ and smaller than $g_\text{1D}\sim -5$ are unachievable. On the attractive side, this is because excited states become bound for $1/g_\text{1D}\to 0^-$, thus moving down through the spectrum. These are seen as the dimmed lines in panel (a). As the states are optimized by minimizing the corresponding energy it is necessary to optimize an infinite number of states to reach $g_\text{1D}=0^-$. This also means that $g_\text{1D}>0$ has to be reached using a repulsive interaction setting an upper limit on $a_\text{3D}$. Instead, $a_\text{3D}$ is controllable through the range of the Gaussian potential though only to a certain extent, as the potential has to be short-range \footnote{For the 1+1 system, $1/g_\text{1D}=0$ was achievable as only one bound state occurred for each resonance.}. The calculations in this chapter are done with $r_0\sim0.1 a_\perp$, i.e. the range is ten times smaller than the transverse length scale $a_\perp=\tfrac{1}{\sqrt{\eta}}\,b$. As the focus will be on repulsive interactions the 2b+1 is shown only for this region in panel (b) of \figref{fig:2f+1_massbalanced_spectrum}. From both panels we see a good correspondence between the two models.   

\begin{figure}[t]
	\centering 
	\begin{tikzpicture}
	\node[] at (0,0){\includegraphics[]{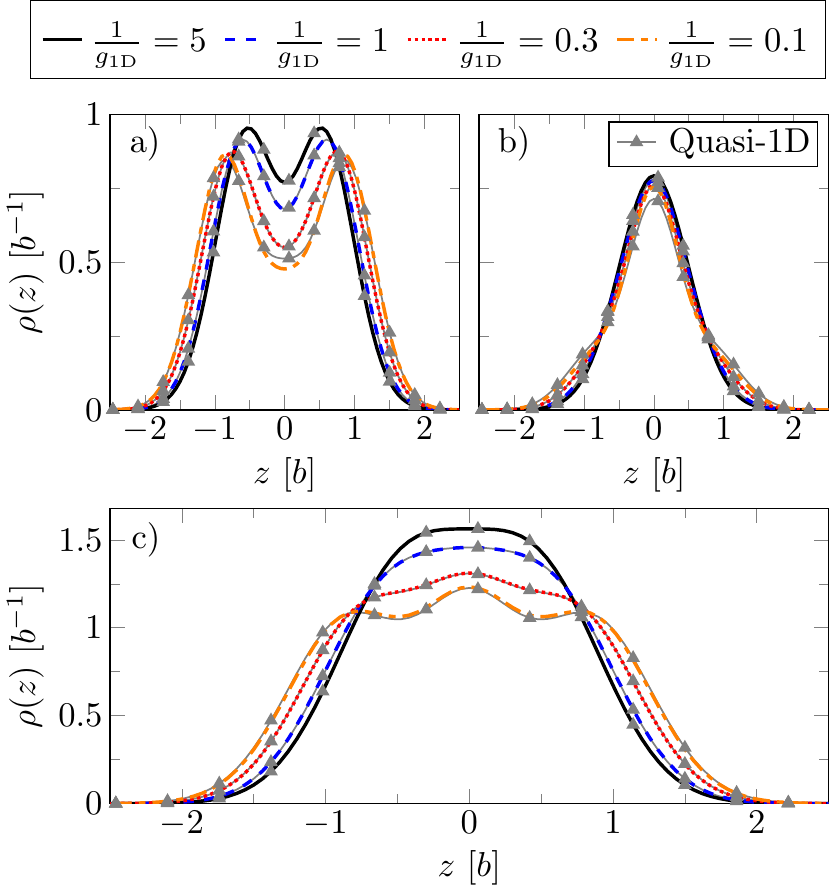}};
	\node[] at (0.05,-6.2){\includegraphics[]{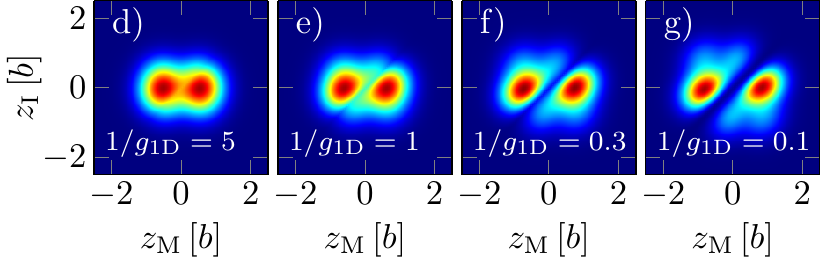}};
	\end{tikzpicture}
	\caption{\label{fig:density_2f+1_mass-balanced_odd}Odd ground state density distribution and pair correlation in the z-axis for the 2f+1 mass-balanced. Lines correspond to the idealized 1D model with zero range interactions and marks are the quasi-1D solution. Panel (a) shows the spin-separated density distribution for the majority particles ($\uparrow$) and panel (b) for the impurity ($\downarrow$). Panel (c) corresponds to the total density. The density distributions are normalized according to the particle number. Density plots in panels (d)-(g) show the pair correlation, for the quasi-1D model, for different interactions, where $z_\text{M}$ ($z_\text{I}$) is the z-position of the majority particles (impurity). The pair correlation is defined in \equref{eq:pair-correlation}.}
\end{figure}

It is interesting to note the horizontal lines in \figref{fig:2f+1_massbalanced_spectrum} panel (a) which are fully antisymmetric non-interacting states. The lowest of these corresponds to having one particle in each of the first three harmonic oscillator states i.e. a fully fermionized state. Such a state is of course not seen in panel (b) where the wave function has to be symmetric under exchange of the identical particles. For $1/g_\text{1D}\to 0$ the two lowest states in panel (a) approach the non-interacting state and become degenerate at the energy of $4\hbar\omega_z$. This is the well-known fermionization limit where interacting particles behave as non-interacting fermions. For a mass-imbalanced system this degeneracy is expected to be lifted as the Hamiltonian no longer fulfills the proper symmetry. Fermionization is also seen in the density distribution plotted in \figref{fig:density_2f+1_mass-balanced_odd} panel (c). It shows the odd ground state for repulsive interactions, where lines are 1D and marks are quasi-1D with $\eta=50$. We see a good correspondence between the two models yet for $1/g_\text{1D}=0.1$ a small deviation is seen in the spin-separated densities. Panels (d)-(g) show the corresponding pair correlation functions in the $z$-axis for the quasi-1D model. The density and pair correlation functions are calculated using \equref{eq:general-form-factor-element} and \appref{app:density-distribution}. For weak interactions, $1/g_\text{1D}=5$, the fermionic nature of the majority particles is seen in both panels (a) and (d). As the interaction becomes more repulsive the majority particles separate in the trap leaving the impurity in the center, as seen in panel (b) and panels (d)-(g). The opposed behaviour was observed for the first excited state where the impurity is located on the edge.

The 1D nature of the system is visible in the pair correlation function, for strong interaction, as the diagonal, i.e. $z_I=z_M$, vanish. This is because the external trap confines the particles to the radial center while the interaction prohibits the particles from being at the same $z$-coordinate.
\begin{figure}[htp]
	\centering 
	\begin{tikzpicture}
	\node[inner sep=0pt] (whitehead) at (0,0)
	{\includegraphics[]{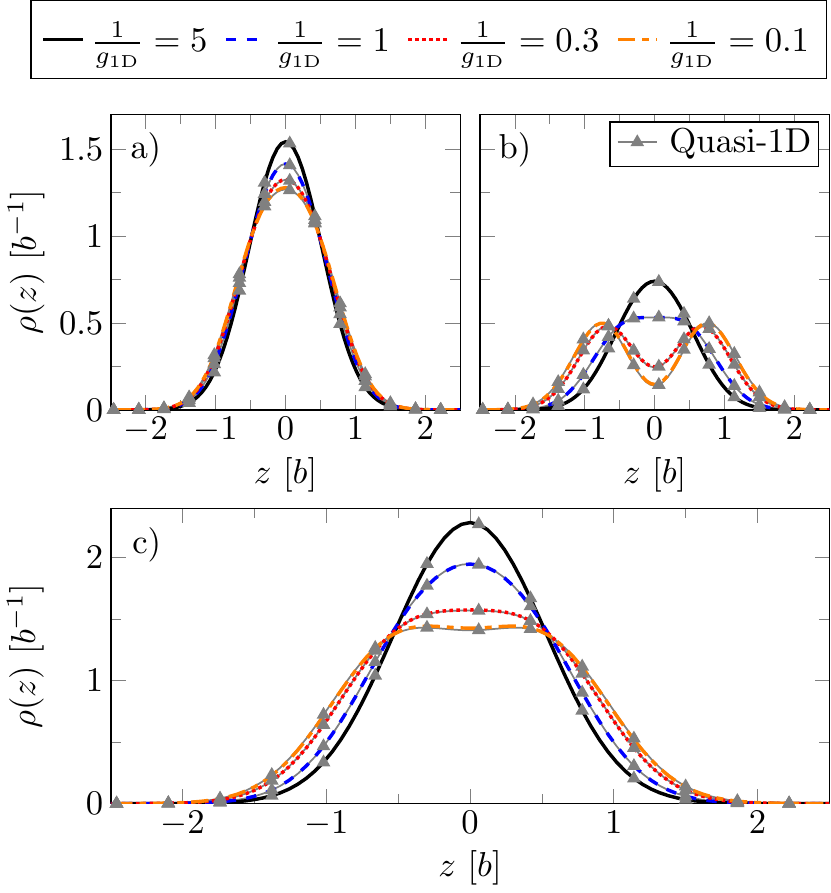}};
	\node[inner sep=0pt](whitehead) at (0.05,-6.2){\includegraphics[]{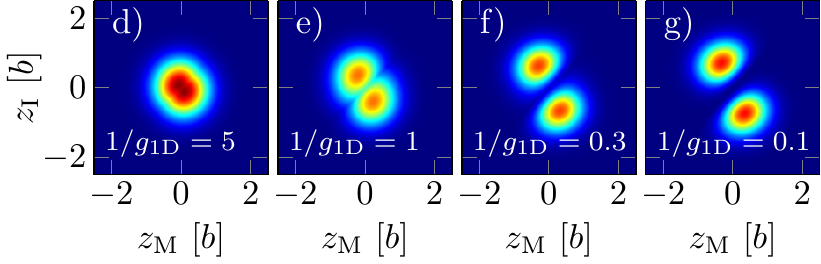}};
	\end{tikzpicture}
	\caption{\label{fig:density_2b+1_mass-balanced_even}Show the same as \figref{fig:density_2f+1_mass-balanced_odd} but for the 2b+1 even ground state.}
\end{figure}

The density and pair correlation functions for the 2b+1 system are seen in \figref{fig:density_2b+1_mass-balanced_even}. Due to the bosonic nature, all particles are located in the center of the trap for weak interaction, i.e. $1/g_\text{1D}=5$. As $1/g_\text{1D}\to 0$, the majority particles stay in the center, as seen in panel (a), while a distinct separation emerges for the impurity in panel (b). This demonstrates that the impurity is located on the edge in the strongly interacting limit, for a mass-balanced system. As for the 2f+1 system, the 1D behaviour is evident in the pair correlation seen in panels (d)-(g).

\begin{figure}[htbp]
	\centering
	\includegraphics{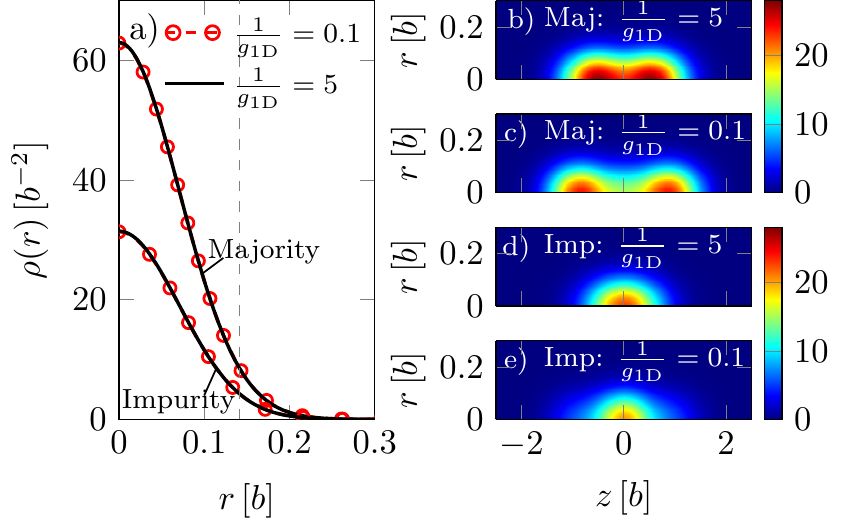}
	\caption{\label{fig:2f+1_strict_1D_radial_dist}Radial distribution for the 2f+1 system with $\eta=50$. Panel (a) show the radial distribution, for two different interaction strength. The dashed vertical line corresponds to the length scale of the transverse trap, i.e. $a_\perp=1/\sqrt{\eta}$ in units of $b$. Notice how no visual difference is seen in the density indicating that the system behaves 1D. Panel (b) shows the axial symmetric density distribution as a function of the radial distance $r$ and the $z$-coordinate for both the majority and impurity. See figure for more information.}
	\vspace{1cm}
	\includegraphics{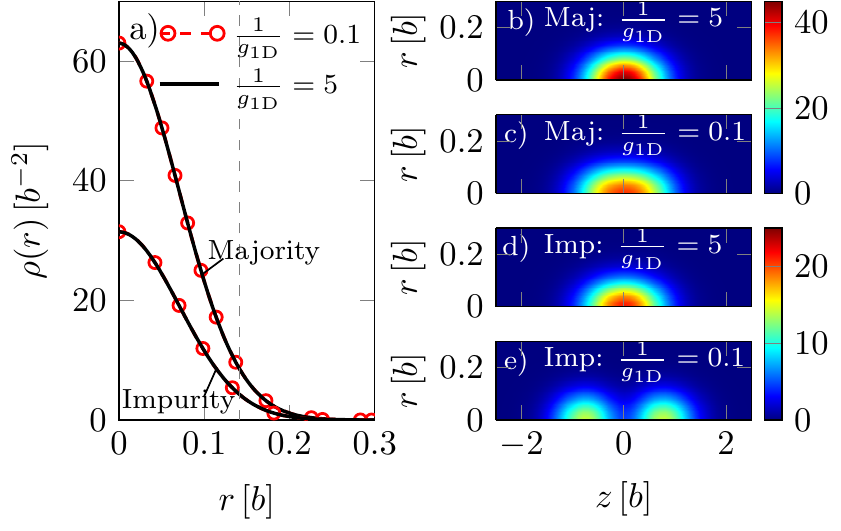}
	\caption{Show the same as \figref{fig:2f+1_strict_1D_radial_dist} but for the 2b+1 system.}
	\label{fig:2b+1_strict_1D_radial_dist}
\end{figure}

As the 2+1 systems have been solved in quasi-1D it is possible to obtain information about the distribution in the radial direction. The radial density distributions for the 2f+1 and 2b+1 are shown in \figref{fig:2f+1_strict_1D_radial_dist} and \ref{fig:2b+1_strict_1D_radial_dist} respectively.

Panel (a) shows the radial density distribution, as a function of the radial distance to the center, $r$, normalized as $N_c=2\pi\int \rho_c(r) r \dd{r}$. The solid lines show the distribution for weak interactions with $1/g_\text{1D}=5$ and the marks for $1/g_\text{1D}=0.1$. Notice how no visible change is seen for the two interactions indicating that the system behaves 1D for both the 2f+1 and 2b+1 systems.

Panels (b)-(e) in \figref{fig:2f+1_strict_1D_radial_dist} show the axial symmetric density distribution as a function of both the $z$- and $r$-coordinated, normalized as $N_c=2\pi\int \rho_c(r,z) r \dd{r}\dd{z}$. Notice the different scales of the radial- and $z$-axis, showing that the system is highly elongated due to the tight trap configuration. Panels (b) and (c) show the distribution of the majority particles for two interaction strength demonstrating how they separates in the trap for large repulsive interactions. Panels (d) and (e) show the corresponding impurity density, where a small deformation for strong interaction compared to the weak is seen.

The 2b+1 system seen in \figref{fig:2f+1_strict_1D_radial_dist} shows a different behavior. Here the majority particles, due to the bosonic nature, are located in the center of the trap. Increasing the interaction, the density decreases toward the center while, again, the spatial extention in the $z$-axis increases. For the impurity, seen in panels (d) and (e), the density vanishes toward the center of the trap, clearly showing that it is located on either side of the majority particles. These feature was also seen in \figref{fig:density_2f+1_mass-balanced_odd} and \ref{fig:density_2b+1_mass-balanced_even} yet here with full 3D density information.

\bibliography{References}

\end{document}